\documentclass{eptcs}
\providecommand{\event}{ADG 2021} % Name of the event you are submitting
\usepackage[T1]{fontenc}
\usepackage{url}
\usepackage{amsmath}
\usepackage{amssymb} 
\usepackage{graphicx}
\usepackage{hyperref}
\usepackage{wasysym}

\newcommand{\comment}[1]{}

\def\orcidID#1{\unskip$^{\mbox{\href{https://orcid.org/#1}{\scriptsize{[#1]}} }}$}

\title{Automated Theorem Proving in the Classroom 
}

\author{Wolfgang Windsteiger\orcidID{0000-0002-7449-8388}
\institute{Research Institute for Symbolic Computation (RISC)\\Johannes Kepler University Linz (JKU) \\
Altenbergerstr. 69, 4040 Linz, Austria \\
\url{Wolfgang.Windsteiger@risc.jku.at}
}
}

\def\titlerunning{Automated Theorem Proving in the Classroom }
\def\authorrunning{Windsteiger}

\begin{document}
\maketitle

\begin{abstract}
We report on several scenarios of using automated theorem proving software in university education. In particular,
we focus on using the Theorema system in a software-enhanced logic-course for students in computer science 
or artificial intelligence. The purpose of using logic-software in our teaching is \emph{not} to teach students
the proper use of a particular piece of software. In contrast, we try to \emph{employ} certain software in order
to spark students' motivation and to support their understanding of logic principles they are supposed to
understand after having passed the course. In a sense, we try to let the software act as a logic-tutor, the 
software is not an additional subject we teach.
\end{abstract}

\section{Introduction}\label{sec:introduction}
In the course of the development of a new curriculum for a bachelor study in computer science at JKU Linz
several years ago we had the opportunity to design a new logic-course~\cite{LOGIC,RISC6097}. 
Logic should play a
more prominent role and will be taught in the first semester for approximately 400 beginner students of computer
science or, since 2019, the then newly introduced bachelor studies in artificial intelligence. The novel
aspects that we wanted to add to a classical course in logic were 

\begin{itemize}
  \item that students should see \emph{logic in action} through the incorporation of logic software and
  \item that certain \emph{modern topics} (such as e.g.~automated or interactive theorem proving and
  satisfiability modulo theories (SMT)) should be 
  presented to the students early in their studies instead of somewhere towards the end in some special lecture.
\end{itemize} 

The course is designed consisting of four more or less independent modules taught by four different lecturers. 
In each
module the lecturer uses the software s*he prefers. In this report we concentrate on module three devoted to 
\emph{reasoning in first-order predicate logic} and supported by the Theorema 
system~\cite{RISC5243,RISC5516}. 

It is important to clarify the content and the didactical goals of the course-module on proving in first-order
logic in order to be able to judge the appropriateness of software support in that lecture. Propositional logic and
reasoning therein as well as the syntax and semantics of first-order predicate logic can be assumed to be familiar 
to the students from previous modules. The ultimate goal of the proving-module 
is to teach students how to do mathematical proofs ``by hand'', because we consider it as an important
capability for computer scientists to being able to reason and argue in a logically sound way, e.g. when talking
about correctness of computer programs.

Unlike in mathematics, where proving is taught to students traditionally through a multitude of proof examples 
exhibited during their studies, we follow the approach of \emph{explicitly teaching the method of proving} via
formal proofs. On the one hand, we present a small set of proof rules and, on the other hand, we consider a
proof to be a finite tree, whose vertices are \emph{proof situations} and each edge represents
\emph{a transition}, i.e.~a logical step, from one situation to another justified by 
one of the above rules. Instead of one of the known classical deduction calculi like sequence calculus, 
Gentzen calculus, or
natural deduction calculus, we use our own set of reasoning rules, which is not necessarily the smallest possible
set that allows a complete reasoning procedure for first-order logic but which aims at generating proofs like
one would do them by hand. Moreover, we present the activity of proving as a \emph{search procedure}
aiming at finding
a successful proof tree, and we explain the transition from a formal proof tree into a traditional 
mathematical proof.

In Section~\ref{sec:theorema} we give a brief overview on the philosophy and the main features of the Theorema 
system. Section~\ref{sec:designLogic} describes the didactical setting how we incorporated Theorema into
the teaching of logic, and, finally, Section~\ref{sec:observations} summarizes some of the insights we gained
from this didactical experiment.
 
\section{The Theorema System}\label{sec:theorema}

\subsection{Theorema: A Brief Overview}

The Theorema system aims to be a computer assistant for the
working mathematician. Support should be given throughout all phases of
mathematical activity, from introducing new mathematical concepts by
definitions or axioms, through first (computational) experiments, the
formulation of theorems, their justification by an exact proof, the
application of a theorem as an algorithm, to the dissemination of the
results in form of a mathematical publication, the build up of bigger
libraries of certified mathematical content, and the like. One focus lies
on the \emph{natural style} of system input (in form of definitions, theorems,
algorithms, etc.), system output (mainly in form
of mathematical proofs), and user interaction. When using the Theorema
system, a user should not have to follow a certain style of mathematics
enforced by the system (e.g. basing all of mathematics on set theory or
certain variants of type theory), rather should the system support the
user in her/his preferred flavor of doing math. 

The development of the Theorema System has been initiated by Bruno Buchberger in the beginning of the 1990's. He
implemented a first version of the Theorema language and some first automated provers in Mathematica. Over the years, the Theorema group at RISC
extended the system remarkably in various directions. Special reasoning methods for elementary analysis, set theory, induction over various domains, geometry, boundary value problems, and many
more have been invented and implemented in the frame of Theorema, see e.g.~\cite{RISC2487,RISC2327,RISC2330}. Since Mathematica~7 released in 2008, the Mathematica Notebook FrontEnd supports
dynamic objects, which allow to implement modern interactive user-interfaces within the Mathematica environment. This was the starting point for Theorema~2.0, a re-design and a complete
re-implementation of the Theorema system with special emphasis on intuitive click-based user interaction, see~\cite{RISC5243}. Theorema~2.0 is open-source and
available through GitHub.

\subsection{Proving with Theorema~2.0}

Theorema~2.0 is implemented as a Mathematica add-on package, and when loaded into Mathematica it presents to the user its main interface component, the \emph{Theorema Commander}.
The Theorema Commander is where the user gets support in various mathematical activities, like doing proofs or computations. Mathematical content can be mixed with text and graphics and is written in \emph{Theorema
notebooks}, which are just Mathematica notebooks using a special stylesheet in order to support special behavior of certain Theorema-specific items, like e.g. definitions or theorems.
When working with Theorema, one creates
a mathematical document using all the capabilities of the Mathematica notebook interface. When one wants to interact with the math, one switches to the Theorema commander in order to initiate 
some action, whose result will then be documented back in the notebook. Hence, Theorema~2.0 can also be seen as \emph{semi-automated mathematical document creator}, which creates parts of a
mathematical document, e.g. proofs, in a fully automated way. We want to emphasize, however, that Theorema is \emph{not just 
an automated prover}, although in this presentation we \emph{concentrate on proving with Theorema}.

In order to generate an automated
proof of some statement, the user would first type the statement into a formula-cell in the notebook, usually
inside
a named Theorem-, Lemma-, or Proposition-environment. All environments may carry names as well as
each formula inside
an environment can carry a separate label as name. Then one would switch to the Theorema commander and choose the 
\textsc{Prove}-activity, which guides through the process of setting up the prover. The proof goal is 
defined by simply selecting the notebook cell containing the goal formula. 
Next is the setup of the knowledge base being available in the proof, which is achieved through the 
\emph{knowledge browser} that shows an outline of each open notebook displaying only formal mathematical content, 
like definitions and theorems, preserving the sectional structure of the notebook while hiding all informal parts, 
such as text and graphics. Sectional groupings can be collapsed
in order to gain overview over the whole document. For the formal entities, the commander does
not display the formulas in detail, it rather shows the formulas' labels only and presents the entire formula as a 
tool-tip when hovering the mouse over the label. Each formula is accompanied with a 
checkbox that, when checked, includes the corresponding formula into the knowledge base.

The next step in the \textsc{Prove}-activity is the \emph{setup of the prover}. A prover in Theorema~2.0 consists 
of individual inference rules that are applied by a proof search engine in a certain order. In the Theorema
commander we are able to activate or deactivate every single inference rule, and we can influence the order, in 
which they are applied, by assigning a priority to each rule. In the final step a summary of all chosen
settings is presented and the prove-task can be submitted. Usually, it
is a good strategy to first run the prover with default settings and a limited search depth and search time (both configurable in the prover setup). In case the proof would not succeed,
the failing proof can be inspected and checked whether certain settings might be changed in order to prevent the prover from running into an undesired path. Otherwise, search depth and search time 
can be increased in order to allow the prover to terminate successfully. When the prover stops it writes an answer back into the notebook right below the environment, from which the proof has been initiated, indicating success
or failure, a version number of the proof, and a link to the automatically generated proof. When clicking the link, a nicely formatted version of the proof explained in natural language is displayed in a separate window, which also offers several options to simplify the proof.

During proof generation and when a proof window is open, the Theorema commander shows a \emph{tree visualization} 
of the proof search. In a successful proof all nodes belonging to a successful branch are colored green, nodes in
failing branches are red, and pending nodes are blue. Pending nodes are proof situations that can still be
handled by one of the available rules. If such nodes are present in a proof this is an indication that the proof
search did not complete, either due to reaching the search time limit or through manual interruption by the user.
Simplification of a successful proof essentially removes all failing branches and pending nodes resulting in an 
all-green proof 
tree that in fact corresponds to a formal proof tree as taught in our logic-course. Click-navigation connects
the Theorema commander and the proof window, i.e.~when clicking a cell in the proof window, an indicator mark 
highlights the corresponding node in the tree view, whereas clicking a node in the tree leads to 
the corresponding textual description of the respective proof step in the proof display. Moreover, when hovering 
the mouse over a node in the tree, the name of the rule applied in that node is displayed as a
tool-tip.

\section{The Design of the Logic Course}\label{sec:designLogic}

\subsection{The Structure of the Course}
\begin{figure}
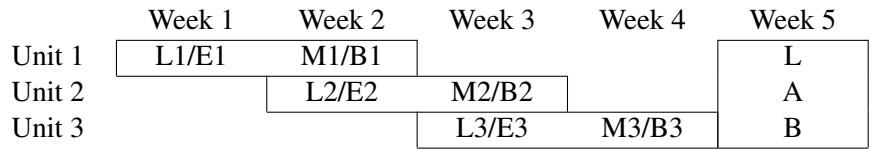

  \begin{center}\setlength{\tabcolsep}{12pt}
    \begin{tabular}{lccccc}
     & Week 1 & Week 2 & Week 3 & Week 4 & Week 5\\ \cline{2-3}\cline{6-6}
    Unit 1 & \multicolumn{1}{|c}{L1/E1} & \multicolumn{1}{c|}{M1/B1} & & & \multicolumn{1}{|c|}{L}\\ \cline{2-4}
    Unit 2 &    & \multicolumn{1}{|c}{L2/E2} & \multicolumn{1}{c|}{M2/B2} & & \multicolumn{1}{|c|}{A}\\ \cline{3-5}
    Unit 3 &  & & \multicolumn{1}{|c}{L3/E3} & \multicolumn{1}{c|}{M3/B3} & \multicolumn{1}{|c|}{B}\\ \cline{4-5}\cline{6-6}
    \end{tabular}
    \caption{Nested Module Schedule}
    \label{fig:scheduleModule}
  \end{center}
\end{figure}

The logic-course is composed of \emph{mandatory} and \emph{elective components} that contribute in different ways 
to the final grade for the whole course. The lecture (L) is accompanied
by an exercise class (E), where students need to solve problems on their own. The predicate logic module consists
of \emph{three units} and the grading of the module is based mainly on the mandatory \emph{minitests} (M)
after each unit.
For each unit there is a voluntary \emph{bonus exercise} (B) and for the whole module there is a voluntary 
\emph{lab exercise}, through which students can enhance a \emph{bad result in a minitest} or
a \emph{bad overall result}.
Of course, students can volunteer to do bonus and lab exercises regardless of whether their achievements 
in the minitests require them to do so. Through publication dates and deadlines for the respective items 
we enforce a nested sequence of lecture/exercise---minitest/bonus for each unit as
depicted in \figurename~\ref{fig:scheduleModule}. This means that in one week we present
the theory unit in the lecture and do concrete examples for that unit in the exercise class, in the following week 
there is the minitest for that unit immediately followed by the bonus exercise and, in parallel, the
presentation plus the exercise class for the next unit.

\subsection{The Use of Theorema in the Frame of the Logic Course}\label{sec:theoremaInLogic}

The features of the Theorema system described in Section~\ref{sec:theorema} make it an ideal companion for
teaching mathematical proving. However, since logic software should not be the main subject
of the course, we decided to move all software-related tasks into the voluntary bonus and lab exercises. 
Other reasons for shifting software aspects into voluntary course tracks include the fact
\begin{itemize}
  \item that some of the software has more the character of a prototypical research software,
  \item that availability and reliability may depend to some extent on operating system requirements,
  \item that, as in the case of Theorema with Mathematica, some software requires other proprietary software that 
  is not available legally free of charge, and
  \item that we have no capacity for in-depth tutorials on how to use the different pieces of software.
\end{itemize}

In each bonus exercise students need to generate one or more proofs from the previous
exercise sheet with the help of the Theorema system, i.e.~they use Theorema to generate an automated proof that 
they previously tried (or succeeded) to do by hand during an earlier exercise. Since Theorema
shows the students both a natural language presentation of the proof as well as
a graphical visualization of the logical structure in form of the proof tree, we consider the Theorema system as
a proof 
tutor. Students can effortlessly investigate how different settings of the prover influence the final proof,
e.g.~what difference it makes whether they allow the prover to apply certain rules or not and what effect it has
whether they apply specific rules earlier than others. Through the proof simplification feature they can observe
how the entire proof search including failing attempts and unused steps finally translates into a proof tree
that corresponds to a successful proof that can then be presented using natural language including all the usual
mathematical phrases as known from their mathematics courses.

The names the Theorema proof tree shows when 
hovering the mouse over the node in the tree correspond to the names of the proof rules introduced during the 
lecture. Again, this should contribute to the students' understanding of how to generate correct and complete 
formal proofs. Over the many years of teaching proving to students of various branches, the main difficulties 
we observed were the following:
\begin{itemize}
  \item At the beginning of a proof, students often have no idea how to start.
  \item They are uncertain, whether particular steps are allowed or not.
  \item They are uncertain, what step to do next.
  \item When the proof is actually finished, they are uncertain whether something else still needs to be shown.
\end{itemize}
The Theorema system just shows them in practice what we try to advertise for
their daily business to remedy the above difficulties: 
\begin{itemize}
\item First write down all formulas in exact syntax and be careful to use the correct syntactical structure.
\item Try to do a formal proof, simplify it, and present it in natural language. 
\item When unsure how to proceed or unsure whether a certain step is allowed or not, concentrate on 
the syntactical structure of the formulas and carefully check, which rules can be applied and which not.
\item Finally, just watch out to close all branches of the tree through application of an appropriate rule.
\end{itemize}

We try to let students experience this procedure in the frame of the lab exercise. When they do the lab exercise 
most of them have done already the three bonus exercises with Theorema, so they are familiar with generating
computer proofs using the system. While in the bonus exercises the requirement was always to submit a 
computer-generated proof, the lab exercise requires them to submit a hand-written proof. To be precise, the
task is to produce a mathematical proof of some simple statement that they know from one of their mathematics 
courses in the style how they do proofs \emph{in these courses}. In order to come up with a proof, the advice is 
to first let Theorema do the proof, then study this proof and try to understand each single step, and finally 
write up their own version of the proof in their own words.

\subsection{Students' Difficulties when Using Theorema}

There are two main issues when working with Theorema, namely \emph{input syntax for mathematical expressions}
and \emph{prover configuration}. We now briefly describe how we try to circumvent them in our use of Theorema in 
the logic-course.

Although Theorema supports \emph{user-friendly two-dimensional input syntax} as used in mathematics textbooks, it 
turns out in practice that people---and this is not limited to students only---have
severe difficulties to enter formulas exhibiting an appropriate syntactical structure using the keyboard. This
is due to the fact that not enough emphasis is put on the tree structure of a formula when teaching the 
language of mathematics. Most people still regard a formula as basically a ``line of (special) characters'', 
inter-mixed with sub- and superscripts and other special constructions such as fractions etc. Theorema offers
\emph{input palettes}, through which the most common formula structures can be entered by mouse-click. 
Provided one proceeds along the tree-structure of the formula, it is guaranteed that the resulting formula
has the right structure because every input button adds invisible parentheses that fix the syntactic structure
regardless of any operator precedences that might be defined in the background. Nevertheless, we hide the
potential difficulties regarding formula input from the students by providing Mathematica notebooks containing
all formulas needed in their exercises. It should be noted, however, that if we had more time available in
the course, it could be worthwhile to use Theorema palette input when teaching the syntax of predicate logic 
in order to 
let students gain practice with the tree structure of mathematical expressions in the language of predicate logic.

Concerning \emph{prover configuration}, Theorema allows the user to turn every proof rule \emph{on} or \emph{off} 
and for every rule to set the \emph{priority}, which influences the sequence in which rules are applied during
the proof search. Of course, Theorema has defined default settings that turn out to generate reasonable proofs
in many cases. Turning off certain rules can lead to ``unprovable theorems'', i.e.~statements that \emph{are} true 
but cannot be proved with Theorema, whereas turning on additional rules will not prevent Theorema from giving
a proof provided it does give one without these rules. What can happen is that it takes longer to find the proof
because additional rules might distract the proof search, or that Theorema generates a different proof when
a new rule guides the proof search into a new direction. If this new direction still gives success then it
may hide another successful proof, because the proof search stops with the first success. The same is true for 
priorities in the configuration, i.e.~assigning different priorities may lead to longer search or to different 
proofs. Moreover, changing the setup may require an adaption of the search depth limit in order to still find
the desired proofs. Playing with these features is an interesting endeavor for advanced students and
professionals, for beginner students it might lead to frustration if the success rate is too low. Therefore,
we provide hints in the notebooks for those cases where the default settings need some adjustment or we provide
estimated computing times for those cases where students should be prepared to wait a little while until the
successful proof appears on the screen.

\section{Observations, Results, and Lessons Learned}\label{sec:observations}

From the arrangement of units in the module as shown in \figurename~\ref{fig:scheduleModule} we see the crucial 
role of 
bonus~1 and~2, because they are enforced to be done \emph{before}
minitest~2 and~3, respectively. From a didactical point of view, this gives us the possibility to design their
content in such a way that students benefit from the bonus for the \emph{next minitest}. Our teaching hypothesis 
is that students improve their own 
proving skills by working with the Theorema system in the ways described in Section~\ref{sec:designLogic}.
Ideally, this improvement should then show an impact in the consecutive minitests.
Any effect from doing the lab exercise cannot be measured through the minitests due to the sequential setup.
However, the content of the lab exercise was targeted towards mathematics-style proving, so doing this exercise
probably contributes to a better overall performance in mathematics.

\subsection{Impact on Logic Minitests}

We compare \emph{average points} in the minitests among groups of students, namely

\begin{description}
  \item[Group All:] all students that attended the minitest.
  \item[Group Bonus~1:] those students who did bonus exercise~1 successfully.
  \item[Group Bonus~1+2:] those students who did \emph{both} bonus~1 \emph{and}~2 successfully.
  \item[Group no Bonus:] those students who did \emph{neither} bonus~1 \emph{nor}~2 successfully.
\end{description}

In our assessments
below, if we claim \emph{statistical significance} of some average being higher or lower than some other then this
is based on a \emph{one-sided T-Test} (assuming different variances) comparing the means of the respective samples.
When we list a $p$-value, then this value is the maximum probability that the averages are in fact equal although
we claim them being unequal. If we omit the $p$-value then it is less than 0.05, i.e.~all our claims below
are statistically safe to more than 95\%. It is important to recall that statistical tests do not reveal
any causalities. This has to be emphasized in particular in our scenario where the groups are not assigned
randomly but students actively join a group or not. Now, if group~A performs better than group~B, this can
be because students are better \emph{because} of being in group~A or the reason can be that the 
\emph{better students}
(more talented, more interested, or more motivated) \emph{choose} group~A\footnote{A superior setup would be 
if we divide the entire class into a group that does the bonus
exercises and compare them to the rest that does not do the bonus exercises, but this is
not feasible in our logic-course because of the voluntary character of the software-related parts.}.

\begin{table}
  \begin{center}
  \caption{Results of minitest~2 (max.~5 points) with $p$-values for equal means. Values in parentheses show
  the size of the groups and column $\diameter$ contains the average scores in the group samples.}
  \label{tab:minitest2}
    \begin{tabular}{l|c||c|c}
    & $\diameter$ & All & Bonus~1 \\ \hline\hline
    All (307) & 3.28 & --- & ---  \\ \hline
    Bonus~1 (139) & 3.62 & 0.002 & --- \\ \hline
    no Bonus (168) & 3.00 & 0.006 & $1.21\times 10^{-6}$
    \end{tabular}
  \end{center}
\end{table}

Summarizing the results of minitest~2 shown in \tablename~\ref{tab:minitest2}, we can say that those who
successfully did bonus~1 perform \emph{significantly better} with an
average score of 3.62 than the overall average with 3.28, whereas those who
did not participate or did not succeed in bonus~1 perform with 3.00 \emph{significantly worse} than both other 
groups. Since the groups are well populated we think that these results are not random. 
\tablename~\ref{tab:minitest3} shows the corresponding results for minitest~3, where the differences are by far 
less accentuated. No group significantly over- or under-performs compared to the overall
average of 3.34, only the average score 3.47 of those who succeeded in both bonus exercises is significantly 
higher ($p=0.04$) than the 3.26 of those who did not participate or did not succeed in both bonus exercises. 
Closest to statistical significance are the $3.42>3.26$ of Group Bonus~1 vs. Group no Bonus ($p=0.08$) and the 
$3.47>3.34$ of Group Bonus~1+2 vs. Group All ($p=0.1$).

\begin{table}
  \begin{center}
  \caption{Results of minitest~3 (max.~5 points) with $p$-values for equal means. Values in parentheses show
  the size of the groups and column $\diameter$ contains the average scores in the group samples.}
  \label{tab:minitest3}
    \begin{tabular}{l|c||c|c|c}
    & $\diameter$ & All & Bonus~1 & Bonus~1+2 \\ \hline\hline
    All (286) & 3.34 & --- & --- & --- \\ \hline
    Bonus~1 (135) & 3.42 & 0.20 & --- & --- \\ \hline
    Bonus~1+2 (104) & 3.47 & 0.10 & 0.33 & --- \\ \hline
    no Bonus (141) & 3.26 & 0.22 & 0.08 & \textbf{0.04}
    \end{tabular}
  \end{center}
\end{table}

\subsection{Impact on Mathematics Skills in General}

We were also interested whether the tutoring by the Theorema system might have a positive effect on the 
mathematical capabilities \emph{in general}, not just on formal proving as taught in the logic-course. We therefore
inspected how students performed in the final exam in the math course ``Discrete Structures'', because, at least
to some extent, the final exam consists of some mathematical argumentation about certain properties in discrete 
mathematics. In this category we compare the overall average (13.56, sample size 166) against those who 
successfully participated in all three bonus exercises (14.73/26), those who did not participate or did not
succeed in any bonus exercise (13.19/88), those who succeeded ($\geq 3$ out of five points) in the lab exercise
(13.70/10), and finally those who succeeded in the lab exercise and in all three bonus exercises (15.00/7). The
only significant relations among those are that those with the bonus exercises are better than those without
and also better than average. One would expect that those with bonus \emph{and} lab would also outperform those
that do worse than the ``bonus only''-students. Statistics does not confirm that due to the small sample size in 
former group with only
7~students, 6~of which score~16 and the last one spoils the result with a score of~9. Without the spoiler this
group would score an average of~16 and be significantly better than \emph{all others}.

\subsection{Self-Assessment of Students when Working with Theorema}

Finally, we also report about some self-assessment from students' side.
As a prerequisite to being able to submit a bonus exercise, students had to answer at most
two standard questions about their experiences with Theorema depending on whether they were successful in 
generating an automated proof with Theorema
(Group~A) or not (Group~B). The possible answers are shown in \figurename~\ref{fig:groupA} and~\ref{fig:groupB}.
For bonus~1--3 there were 274, 251, and 180~self-assessments, respectively, and
the ratio A:B was a constant 2:1. Again, with these high numbers we consider the results to be non-random.

\begin{figure}
\fbox{
\begin{minipage}{0.97\textwidth}
\footnotesize
\begin{enumerate}
  \item I did not try or was not able to do the examples by hand, but now I think would be able to do them.
\item I did not try or was not able to do the examples by hand. I think I would still not be able to do such proofs.
\item I had no problems doing the proofs by hand. However, they are different from the Theorema proofs and I'm confused now whether my proofs are wrong.
\item I had no problems doing the proofs by hand. However, they are slightly different from the Theorema proofs because Theorema uses certain rules that I did not know. Still, I think my proofs are fine.
\item I had no problems doing the proofs by hand. However, they are slightly different from the Theorema proofs and in the future I would do my proofs differently.
\item I had no problems doing the proofs by hand. After doing the proofs with Theorema I realized that at least one of my original proofs was wrong.
\item I had a hard time doing the proofs by hand. However, I think when doing the next proof by hand, it will be equally difficult, doing the proof with Theorema did not help me for improving my own skills.
\item I had a hard time doing the proofs by hand. After doing the proof with Theorema I understand much better how all of this works. I feel that my own skills improved by using Theorema.
\item I don't see any connection between the examples from the exercises and the Bonus Exercise with Theorema
\end{enumerate}
\end{minipage}}
    \caption{Possible answers for Group~A.}
    \label{fig:groupA}
\end{figure}

\begin{figure}
\fbox{
\begin{minipage}{0.97\textwidth}
\footnotesize
\begin{enumerate}\setcounter{enumi}{9}
  \item I did not try or was not able to do these examples by hand. I wanted to see how Theorema does the proofs, but I failed to produce a compete proof.
\item I did not try or was not able to do these examples by hand. Theorema is much too complicated for me to use it for such exercises.
\item I had no problems doing the proofs by hand. Unfortunately, I failed to produce a complete proof with Theorema. It would have been interesting to compare.
\item I had no problems doing the proofs by hand. I'm not interested how an automated proof looks, I have done them by hand anyway.
\item I had a hard time doing the proofs by hand. Unfortunately, I failed to produce a complete proof with Theorema. It would have been interesting to compare.
\item I had a hard time doing the proofs by hand. I'm not interested how an automated proof looks, I have done them by hand anyway.
\item I don't see any connection between the examples from the exercises and the Bonus Exercise with Theorema.
\end{enumerate}
\end{minipage}}
    \caption{Possible answers for Group~B.}
    \label{fig:groupB}
\end{figure}

First we analyze Group~A. In bonus~1 and~2 as well as when we take all three bonus exercises together there
is a rather significant gap between the top-four answers 
(4--1--5--8), which sum up to~\textasciitilde 75\%, and
the trailing five with~\textasciitilde 25\%. In bonus~3 the top-four are still the same and make up 72\%, but now 
the top-two are~1 and~8 and make up 42\%, which had severe problems with proving by hand but consider
the Theorema-tutoring to be of help. Still, we want to mention that answer~7 (``proving stays equally
difficult even after Theorema-tutoring'') almost made it into the top-four
in bonus~3, which can probably be explained through the fact that exercise and bonus~3 were the most difficult
with quantifier proving, proper use of definitions, and induction in the natural numbers, the latter not being
supported by Theorema. The most stunning result is, however, Group~A.8 (``had a hard time doing
proofs by hand but feel they had improved through Theorema-tutoring''), which showed one of the weakest 
performances
in minitest~2 with an average of 2.82 points (rank~14) and improved to rank~4 in minitest~3 with an average of
3.47 and being second-biggest group in bonus~3. Also interesting, but less easy to explain is Group~A.9,
i.e.~those who succeeded with Theorema but do not see any connection to doing their proofs by hand. They improve
from rank~8 in minitest~2 with an average of 3.00 to rank~1 in minitest~3 with an average of 3.86, where in
almost half of the comparisons to other groups the better average is even statistically significant.

The picture is far less accentuated in Group~B, where answer~13 is the most popular in bonus~1, bonus~3, and also 
over all three exercises with~\textasciitilde 25\%. Answer~12 starts in second place with~\textasciitilde 23\% in
bonus~1, is then by far the most popular choice in bonus~2 with~\textasciitilde 35\% and falls down to sixth
place in bonus~3, only trailed by answer~15. Those that do not see any connection between the exercises and the
bonus are fortunately always the smallest portion, both in Group~A and~B, except for bonus~3 in Group~B.
An interesting development is displayed by Group~B.14, which performs second-weakest with an average of~2.51
in minitest~2 and improves to second-best with an average of~3.58 in minitest~3, so maybe just spending the
time with Theorema can spark the interest and improve performance even if the proof with Theorema fails.

Analyzing Groups~A and~B together we see that the top-four of Group~A stay top-four in the same order overall
(4--1--5--8) followed by the top-two of Group~B (12 and 13). It should be mentioned, however, that those, who
were not able to do proofs by hand and believed they could do them after Theorema-tutoring (Group~A.1, rank~2) 
could not justify their bold claim in the minitests, where they only rank $10^{\text{th}}$ by their average scores.

\section{Conclusion}

We report on a classroom experiment using the automated theorem proving software Theorema in the teaching of logic.
We describe how software is applied to aid the learning process of students, how tutoring-by-software 
correlates with students' performance in exams, and we report on how students experienced their being
tutored by software. Some interresting connections between classroom-use of software and students' own
proving capabilities were observed, most notably that those who had a hard time doing proofs by hand 
in the frame of the exercises and claimed an improvement of their understanding through being tutored by software 
showed a significant improvement from one exam to the next.

\bibliographystyle{eptcs}
% \bibliography{refs,ww-own}

\begin{thebibliography}{1}

\providecommand{\url}[1]{\texttt{#1}}
\providecommand{\href}[2]{\texttt{#2}}
\providecommand{\urlalt}[2]{\href{#1}{#2}}
\providecommand{\doi}[1]{doi:\urlalt{http://dx.doi.org/#1}{#1}}
\providecommand{\eprint}[1]{arXiv:\urlalt{https://arxiv.org/abs/#1}{#1}}
\providecommand{\bibinfo}[2]{#2}

\bibitem{LOGIC}
A.~Biere, W.~Schreiner, M.~Seidl, and W.~Windsteiger.
\newblock {Logic for Computer Science}, 2020.
\newblock Course in the first year in the bachelor program for computer science
  at Johannes Kepler University Linz (JKU), taught since 2013.

\bibitem{RISC2327}
B.~Buchberger.
\newblock {Theorema: A Proving System Based on Mathematica}.
\newblock {\em The Mathematica Journal}, 8(2):247--252, 2001.

\bibitem{RISC2487}
B.~Buchberger, A.~Craciun, T.~Jebelean, L.~Kovacs, T.~Kutsia, K.~Nakagawa,
  F.~Piroi, N.~Popov, J.~Robu, M.~Rosenkranz, and W.~Windsteiger.
\newblock {Theorema: Towards Computer-Aided Mathematical Theory Exploration}.
\newblock {\em Journal of Applied Logic}, 4(4):470--504, 2006.
\doi{10.1016/j.jal.2005.10.006},

\bibitem{RISC2330}
B.~Buchberger, C.~Dupre, T.~Jebelean, F.~Kriftner, K.~Nakagawa, D.~Vasaru, and
  W.~Windsteiger.
\newblock {The Theorema Project: A Progress Report}.
\newblock In M.~Kerber and M.~Kohlhase, editors, {\em {Symbolic Computation and
  Automated Reasoning (Proceedings of CALCULEMUS 2000, Symposium on the
  Integration of Symbolic Computation and Mechanized Reasoning)}}, pages
  98--113. St. Andrews, Scotland, Copyright: A.K. Peters, Natick,
  Massachusetts, 6-7 August 2000.

\bibitem{RISC5243}
B.~Buchberger, T.~Jebelean, T.~Kutsia, A.~Maletzky, and W.~Windsteiger.
\newblock {Theorema 2.0: Computer-Assisted Natural-Style Mathematics}.
\newblock {\em JFR}, 9(1):149--185, 2016.
\doi{10.6092/issn.1972-5787/4568}. 

\bibitem{RISC6097}
D.~M. Cerna, M.~Seidl, W.~Schreiner, W.~Windsteiger, and A.~Biere.
\newblock {Computational Logic in the First Semester of Computer Science: An
  Experience Report}.
\newblock {\em {CSEDU 2020}}, pages 374--381, 2020.
\doi{10.5220/0009464403740381}.

\bibitem{RISC5516}
W.~Windsteiger.
\newblock {Theorema 2.0: A Brief Tutorial}.
\newblock In T.~Jebelean and D.~Zaharie, editors, {\em {Proceedings of SYNASC
  2017}}, IEEE Explore, pages 1--3, 2017.
\doi{10.1109/SYNASC.2017.00016}.
\end{thebibliography}

\end{document}